\documentclass[10pt, conference, compsocconf]{IEEEtran}
\ifCLASSINFOpdf
\else
  \usepackage{graphicx}
\fi
\usepackage{url}
\begin{document}
\title{Airborne software tests on a fully virtual platform}


\author{\IEEEauthorblockN{Famantanantsoa Randimbivololona\IEEEauthorrefmark{1},
 Abderrahmane Brahmi\IEEEauthorrefmark{2} and Philippe Le Meur\IEEEauthorrefmark{3}}
\IEEEauthorblockA{\IEEEauthorrefmark{1}Avionics and Simulation Products\\
Airbus Operations SAS\\
Toulouse, France\\
Email: famantanantsoa.randimbivololona@airbus.com}
\IEEEauthorblockA{\IEEEauthorrefmark{2}Avionics and Simulation Products\\
Airbus Operations SAS\\
Toulouse, France\\
Email: abderrahmane.brahmi@airbus.com}
\IEEEauthorblockA{\IEEEauthorrefmark{3}Avionics and Simulation Products\\
Airbus Operations SAS\\
Toulouse, France\\
Email: philippe.le-meur@airbus.com}
}


%Telephone: (800) 555--1212, Fax: (888) 555--1212}


\maketitle

\begin{abstract}
This paper presents the early deployment of a fully virtual platform to perform the tests of certified airborne software.
This is an alternative to the current approach based on the use of dedicated hardware platforms.
\end{abstract}


% \fi
\IEEEpeerreviewmaketitle

\section{Introduction}
The verification process is a major process in the development of airborne software. Its purpose is to detect and report errors
that may have been introduced during the development. It is typically a combination of reviews, analyses and test.
The means used to satisfy the verification objectives have to be checked as being technically correct and
 complete for the software level\cite{DO178B}.
Testing is assigned two complementary objectives\cite{DO178B}:
\begin{itemize}
\item One objective is to demonstrate that the software satisfies its requirements.
\item The second one is to demonstrate with a high degree of confidence that errors which could lead to unacceptable failure conditions,
as determined by the system safety assessment process, have been removed.
\end{itemize}

Currently, tests of software are performed on dedicated hardware means: execution platforms, analyzers, integration benches. The limitations incurred while using such dedicated means 
are known since a while: cost, obsolescence, lacking of debug capability, intrusive test technique,
 lacking of observability. The latter two are noticeably more acute when dealing with the tests of dependability-related properties. 
Such tests require specific conditionings of the state of the hardware means and/or some specific software context settings. 
This is not always technically achievable, nor practically affordable in a non intrusive manner.

The use of a fully virtual platform is an alternative approach able to yield effective solutions to those limitations,
provided it can take into account the requirements assigned to the test means of airborne software.
 
\section{SIMUGENE: A test-focused virtual framework}
\subsection{Technology overview}
The SIMUGENE virtualization framework has been setup to address the needs of airborne software.
It is not a simulator, it is a simulator generator. It is equipped with a full modelling and development environment, systemDDK, 
whose generator produces automatically upto eighty percent of the code (C++) of a simulator.
For the processor model, it uses a bit-accurate instruction set simulator (ISS): a SIMUGENE simulator runs unmodified binary code.
The interpretation technique has been retained for the ISS. There is no use of dynamic translation from the target binary towards a host binary.
All the ISS are provided in an external library.

The operational schema to develop -and later run- a virtual platform using the SIMUGENE framework is as depicted below:\\

\begin{figure}[h]
\centering
\includegraphics[width=2.0in]{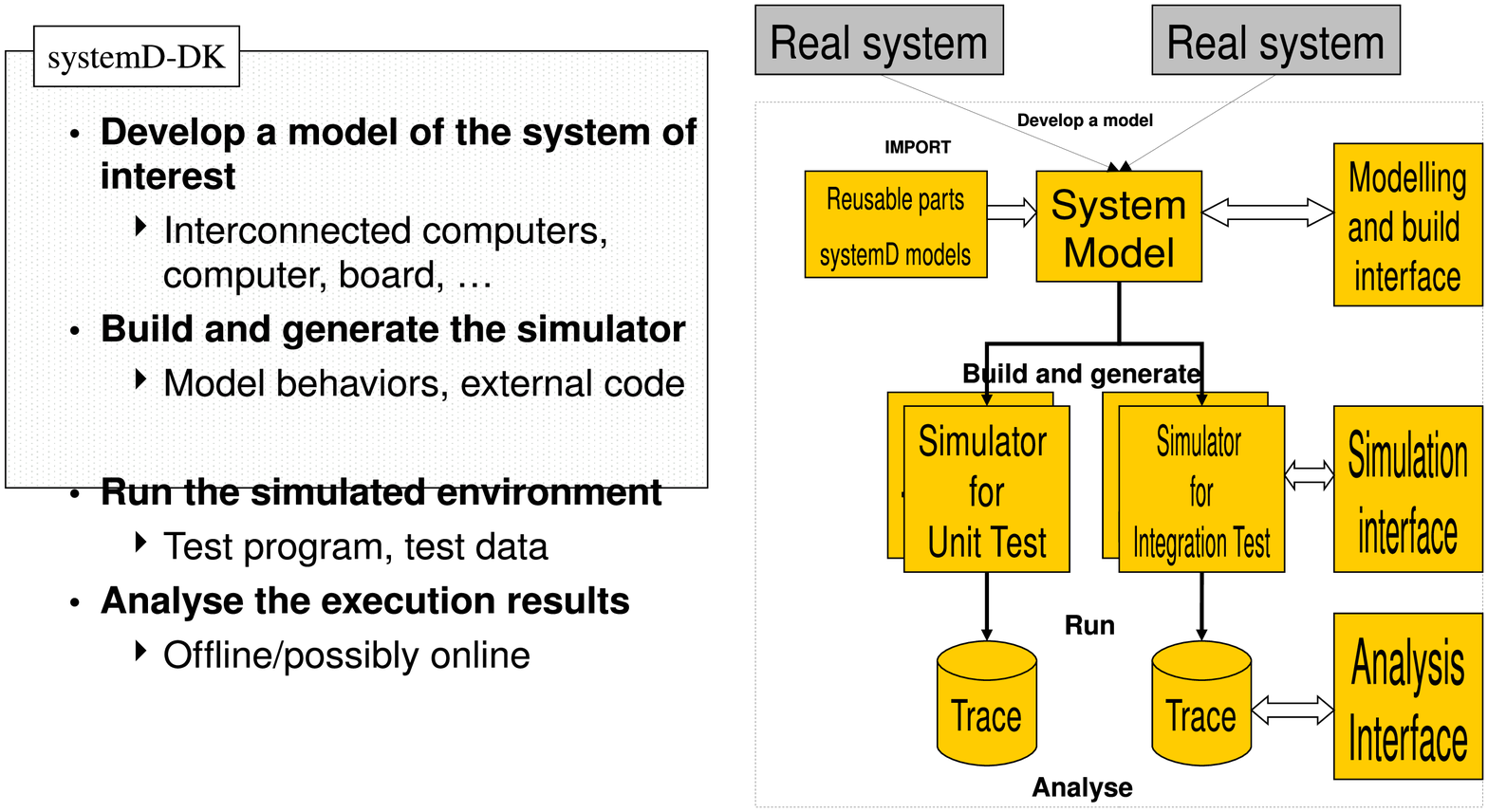}
% where an .eps filename suffix will be assumed under latex, 
% and a .pdf suffix will be assumed for pdflatex; or what has been declared
% via \DeclareGraphicsExtensions.
\caption{SIMUGENE operational schema.}
\label{fig:SIMUGENE operational schema}
\end{figure}

All the phases of this schema (modelling, development, verification, validation) as well as the technical choices are mastered 
in order to ensure the conformance (as a test means) and the representativity (scope, limit of use) of each virtual platform.

\section{Scope of use in the software verification process}
\textbf{Test types coverage.}

A virtual platform may cover a large perimeter of tests:
\begin{itemize}
\item Low level testing: To verify the implementation of software low-level requirements into unit components.
\item Software integration testing: To verify the interrelationships between software components;
to verify the implementation of the software components within the software architecture.
\item Hardware/Software integration testing: To verify correct interfacing and control of the hardware by the software.
\item Final integration testing: To verify correct operation of the whole software in the final target computer environment.
\end{itemize}

The final hardware / software integration tests must be performed using the real hardware.  These are mainly performance tests and hardware control tests.
The targeted test framework is an optimized combination of virtual platform and real hardware computer
(e.g. ninety percent on virtual platform have been achieved on the early deployments).

\textbf{Virtual fault injection for robustness testing.}

Robustness testing is an integral part of the verification of an airborne software.
They can take place in any test phase and their implementation is not simple while using the current dedicated hardware platform.
E.g. How to condition the test to verify that the software behave correctly when writing into an E2PROM whose response time lasts between 3 ms and 10 ms ?
The problem here is that the response time of the currently used E2PROM component cannot be changed.
Commonly a dedicated piece of software is added to inject controlled faults but it is an intrusive technique for the software under test.
\vfill\break

Within a virtual platform, each component can be extended to embed its faulty behaviors and any transaction within the platform can be altered by a faulty behavior from its environment.
These internal and external faults are controlled in the same way within a SIMUGENE simulator (start, stop, frequence, type) and they can be combined as required.
This virtual fault injection technique is non intrusive for the software under test and it has been applied at different test types ranging from low level test upto the final integration test. 

The figure below shows an overview of a virtual computer augmented with a virtual fault injection model.

\begin{figure}[h]
\centering
\includegraphics[width=2.0in]{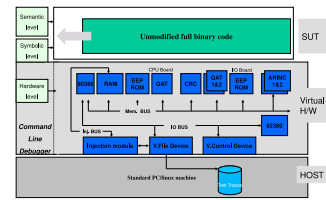}
% where an .eps filename suffix will be assumed under latex, 
% and a .pdf suffix will be assumed for pdflatex; or what has been declared
% via \DeclareGraphicsExtensions.
\caption{SIMUGENE full system.}
\label{fig:SIMUGENE full system}
\end{figure}

\section{Conclusion}
The introduction of virtual platforms provides effective solutions to recurring problems on the test means.
One of the most noticeable solution is the non-intrusive virtual fault injection technique which still can be largely improved.

%\section*{Acknowledgment}

%more thanks here

% FR: 2012 Cutting point control for the equalization of the last page
%\vfill\break

%\IEEEtriggeratref{1}


%\bibliographystyle{IEEEtran}
% FR:usage => \begin{thebibliography}{number of bibitems}

\end{document}